\documentclass[sigconf]{acmart}

\setcopyright{none}
\renewcommand\footnotetextcopyrightpermission[1]{}

\usepackage{url}
\usepackage{array}
\usepackage{soul}
\usepackage{amsmath}
\usepackage{amsthm}
\usepackage{xcolor}
\usepackage{algorithm}
\usepackage[noend]{algpseudocode}
\acmConference[]{}{}{}

\usepackage{fancyhdr}
\def\refname{}

\usepackage{balance}

\begin{document}

\title{Digital Contact Tracing: Technologies, Shortcomings, and the Path Forward}

\author{Amee Trivedi}
\affiliation{
	\institution{University of Massachusetts, Amherst}
}
\email{amee@cs.umass.edu}

\author{Deepak Vasisht}
\affiliation{
	\institution{Microsoft \& University of Illinois Urbana-Champaign}
}
\email{deepakv@illinois.edu}

\begin{abstract}
Since the start of the COVID-19 pandemic, technology enthusiasts have pushed for digital contact tracing as a critical tool for breaking the COVID-19 transmission chains. Motivated by this push, many countries and companies have created apps that enable digital contact tracing with the goal to identify the chain of transmission from an infected individual to others and enable early quarantine. Digital contact tracing applications like AarogyaSetu in India, TraceTogether in Singapore, SwissCovid in Switzerland, and others have been downloaded hundreds of millions of times. Yet, this technology hasn't seen the impact that we envisioned at the start of the pandemic. Some countries have rolled back their apps, while others have seen low adoption \cite{techreviewsuccess, techreviewtracker}. 

Therefore, it is prudent to ask what the technology landscape of contact-tracing looks like and what are the missing pieces. We attempt to undertake this task in this paper. We present a high-level review of technologies underlying digital contact tracing, a set of metrics that are important while evaluating different contact tracing technologies, and evaluate where the different technologies stand today on this set of metrics. Our hope is two fold: (a) Future designers of contact tracing applications can use this review paper to understand the technology landscape, and (b) Researchers can identify and solve the missing pieces of this puzzle, so that we are ready to face the rest of the COVID-19 pandemic and any future pandemics. A majority of this discussion is focused on the ability to identify contact between individuals. The questions of ethics, privacy, and security of such contact tracing are briefly mentioned but not discussed in detail.
\end{abstract}




\maketitle

\section{Introduction}
\label{sec:introduction}


COVID-19 has affected 33,349,007 lives worldwide causing more than 1,003,010 deaths globally \cite{worldometercovid}, impacting the mental health of citizens, crashing economies, and creating global financial distress, thus giving rise to the start of a global recession. The disease burden of infectious diseases such as COVID-19, SARS, pandemic influenza, etc. is very high and contact tracing is a critical step in the containment and reduction of infectious disease spread. Contact tracing is a method of identifying and testing all people who have come in contact with an infected person and iterating through the process for each of the contacts. Contact tracing helps in greatly reducing the spread of disease by identification, and isolation of possible spreaders while also provisioning services for counseling or diagnosis of the illness.



The spread of communicable diseases such as measles, coronavirus (COVID-19), SARS as well as sexually transmitted diseases can be restricted significantly using contact tracing. Currently, once an individual test positive for a communicable disease, the public health workers question the identified patient about the recent history of their social interaction to create a \emph{contact list} comprising of people who came in contact with the infected patient. As a next step, the public health agency further examines and tests each identified person in the contact list. If anyone among the contacts tests positive, that person is labeled infected, isolated, and the process is repeated. This process of tracing is a very labor-intensive process. Additionally, it is very hard for individuals to remember all their social contacts in a particular timeframe resulting in incomplete and inaccurate information. Moreover, most pathogens stay active at a location even after the patient might have left a location. So if a patient has spread the pathogens of an infectious disease (by coughing, touching surfaces with infected hands, etc) at a location, the individuals who visit the same location after the patient's departure while the pathogens are still active might be at risk of getting infected. Such cases are extremely hard to identify since they fall beyond the social interactions of the patient.

\begin{figure}
\begin{centering}
\includegraphics[width=0.75\columnwidth]{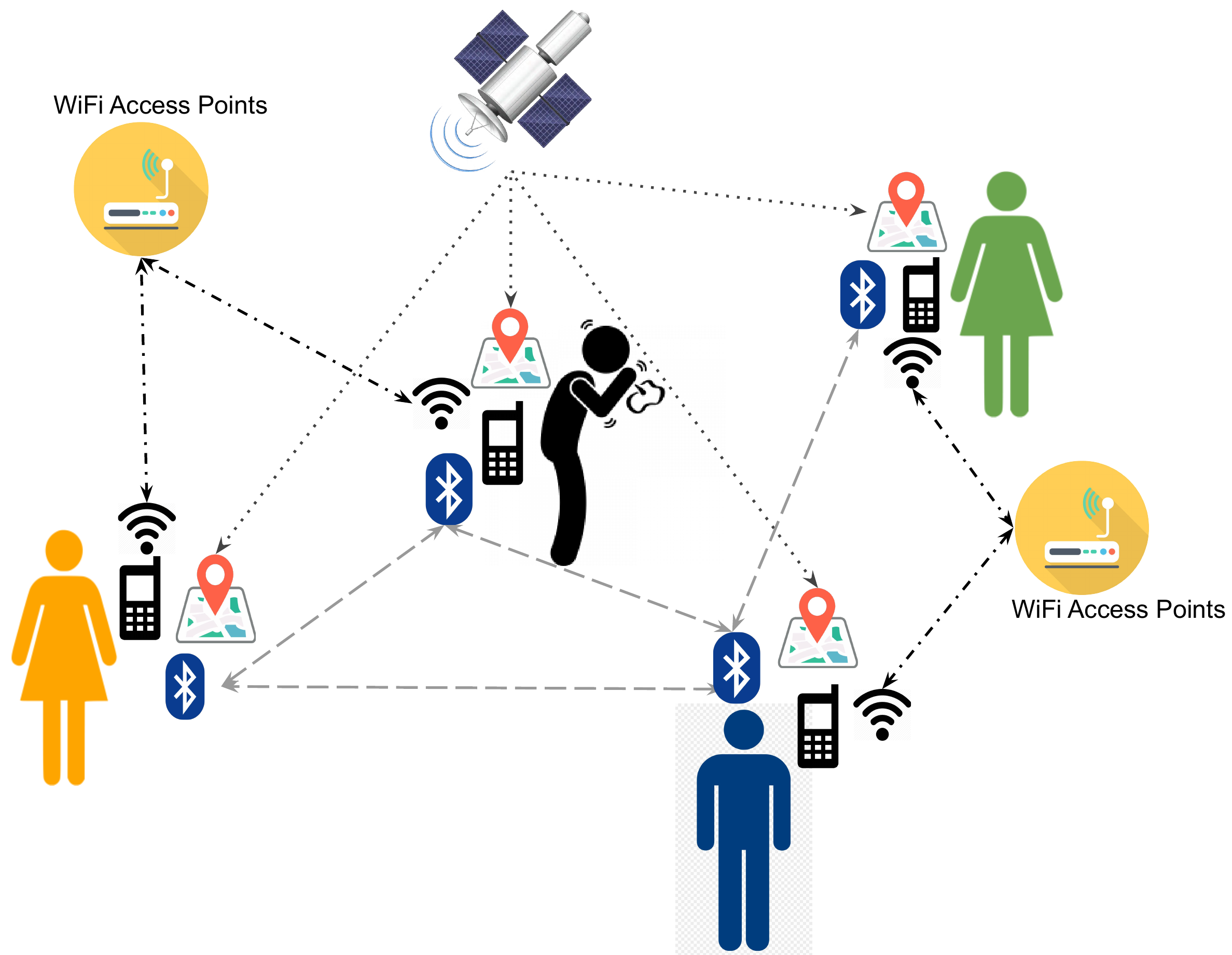}
\vspace{-0.1in}
\caption{Digital contact tracing technologies can help trace contacts between infected individuals and others.}
\label{fig:tech}
\vspace{-0.3in}
\end{centering}
\end{figure}
The spread of infectious disease is fundamentally linked to the potential of transmission of the pathogens over face-to-face interactions or in-person contact with infected physical objects (such as doorknobs, table surfaces, etc). These interactions or connections that provide a mode for disease propagation define a network that provides an insight in understanding the potential structure of transmission that can be used for contact tracing. Additionally, there are many different factors such as pathogen active period,  host immunity, duration of contact, the volume of pathogens, etc. that impact the possible routes of transmission. 

While there has been significant interest in technology-aided tracing based using Bluetooth \cite{aarogya_setu, trace-together, SwissCovid, COVIDSAFE}, QR code \cite{yasaka2020peer}, GPS \cite{CovTracer}, and WiFi traces \cite{trivedi2020wifitrace, TraceFi} as shown in fig \ref{fig:tech}, each technology has its benefits and limitations. In this work, we try and provide an overview of all the technologies, their limitations, questions to ponder on while selecting and designing a digital contact tracing system, and a brief discussion on the path forward.

\section{Digital Contact Tracing}

Digital contact tracing is the method of identification and determination of contact between 2 or more users using a technology-enabled tracking system. In the context of COVID-19, contact is defined as being within 6 feet of an individual for a period of over ten minutes. Manual contact tracing is time-consuming and requires a lot of manual labor, which when coupled with a poor human recollection of social encounters in the recent past creates a possibility of incomplete and inefficient contact tracing. These limitations can be overcome with digital contact tracing systems. The main advantages of a digital contact tracing solution are (i) it creates a log of locations visited (absolute or relative locations) and time of visit (ii) passive logging requires no human intervention (iii) scalability to a large audience, and (iv) provision for easy and quick notification interface. 

The primary ubiquitous tool for digital contact tracing is the mobile phone. In today's world, users carry their phones everywhere and mobility trajectories of mobile phones act as a proxy for user/owner mobility. Hence, using mobile phones for implementing contact tracing systems seems quite reasonable. Based on the various features present in a smartphone, the digital contact tracing systems can be designed based on GPS, Bluetooth, WiFi, etc. with each technology offering its share of benefits and limitations. In this section, we start by providing an overview of metrics to measure a digital contact tracing system followed by a description, pros, and cons of the available technology-based contact tracing solutions while keeping the metrics in mind. We would like to emphasize that any digital contact tracing solution is complementary to the manual contact tracing solutions and is not designed to replace manual contact tracing. 

\subsection{Metrics}
\label{sec:metrics}
In this section, we identify the key metrics for measuring the effectiveness of a contact tracing system:

\vspace{6pt}\noindent\textbf{Accuracy: }
First, a digital contact tracing solution should have high spatial and temporal accuracy to correctly identify proximity and co-location. The accuracy of users in proximity as identified by the tool v/s observed ground truth is heavily tied to the location granularity of the tracking system. A contact tracing system that uses Call Data Records (CDR) datasets with a coarse location granularity in the range of Kms will have very low accuracy (high false positives) as compared to a Bluetooth based tool with a fine location granularity in the range of few meters. Typically, we objectively measure accuracy by computing the false positives and false negatives of the system. In this context, false positives identify the number of people falsely identified as being in contact, i.e. people who were never in contact with a COVID-positive patient but were identified as a contact. While false negatives identify the people who were actually in contact with a COVID-positive patient but were not identified as one. Both false positives as well as false negatives are important. On one hand, if the number of false positives is high, the contact tracing system will identify a lot more people in need of tests causing false alarm, reducing trust in the application, and overwhelming the healthcare resources. On the other hand, a high false negative value indicates that the system is missing out on identifying several colocated users thereby increasing the risk to the general population. Technology users view accuracy as very important factor for their willingness to adopt such solutions. According to a recent survey by Microsoft Research \cite{MSRStudy}, around 85\% of the surveyed population was willing to install contact tracing apps that were perfectly accurate.

\vspace{6pt}\noindent\textbf{Privacy: }
Any contact tracing solution needs to interact with user health and location data, which makes privacy a very important metric. In fact, in the Microsoft research study quoted above \cite{MSRStudy}, privacy concerns were almost as high on people's minds as accuracy. Systems that collect absolute location information could leak details about user mobility revealing entire trajectory with details about absolute location visit history, frequency of visit, and details of frequent associations over time. 
Alternatively, systems that collect relative information could leak frequent association information as well as mobility patterns. Additionally, with a lot of social content available, and people self-reporting check-ins on social media, merging multiple datasets could result in a lot more privacy leakage. Understanding the system areas where privacy can be compromised and mitigating the risks of compromise is an important measure.

\vspace{6pt}\noindent\textbf{Data Security and Ease of Attacks: }
Designing a contact tracing application requires the designers to make decisions about the amount of data collected per user and about what data is being stored. Is it a single central data storage or is it a distributed data store? How easy and susceptible is the data store to attacks?

\vspace{6pt}\noindent\textbf{Ubiquity: }
For a contact tracing system to be effective, a large percentage of the population must adopt it. Accordingly to \cite{WSJ-TT}, about 80\% of the user population needs to use the contact tracing app for it to be effective. Such large scale user adoption needs:  
\begin{itemize}
\item[(a)] Compatibility: 
The technology used for proximity detection must be compatible with smartphones using outdated software. For eg., 25\% of global Android users would be unable to use apps designed below Android 8.0 that was launched in 2017. Therefore, any apps that use capabilities available only in modern mobile operating systems won't be compatible with large user devices resulting in a selective representation of the population. 
\item[(b)] Scalable design: The app must be able to identify contact not only in homes where few smartphones talk to each other and identify their distance, but also in crowded retail stores, subways, and supermarkets, that have hundreds of people walking around. The design of the apps must be able to handle such large traffic and measure accurate proximity within this large group of people. The other aspect of the scalable design is the ability to not crash when tens of millions of people end up using the app -- this is the scale that most internet applications dream of. For example, AarogyaSetu, the official Indian government application for digital contact tracing crossed 100 million downloads. Apps operating at such a scale need to follow skillful engineering practices.
\end{itemize}

This challenge is amplified because because not only do the users need to download, install the app, and consent the collection of data but also, in some cases (e.g. Bluetooth-based apps) to keep the app running and enabled all the time. Such apps may drain energy very quickly resulting in users disabling the apps. This on a bigger scale results in an inefficient and incorrect contact tracing system. From the user adoption perspective, systems that use an existing data source collected passively for alternate reasons such as WiFi logs based system that uses network logs collected for network analysis, do not suffer from this problem. Once a user gives consent to data collection the network-centric system does not need any apps to run on individual mobile devices.

\subsection{Data Sources and Technology}
In this section, we describe various technologies used for contact tracing, their advantages, and their limitations.

\vspace{6pt}\noindent\textbf{Bluetooth - }Bluetooth is a protocol for short-distance wireless communication (up to 100 meters) using radio waves. For two devices to communicate using Bluetooth, they should be equipped with a Bluetooth chip and have Bluetooth communication enabled. The ubiquity of smartphones along with Bluetooth availability on a large range of smartphones (and even some feature phones) has made it a very promising candidate for digital contact tracing. 

Applications (Apps) designed to use only Bluetooth technology identify the presence of other Bluetooth devices within their proximity range (~100 meters) as contacts. So, in a densely populated location such as a large retail store or a residential housing complex with closely knit small apartment spaces, everyone within the radius of 100 meters of a Bluetooth device might be marked as a contact in close proximity. Though the person might be in a different shop or apartment with no physical contact. To alleviate this concern, Bluetooth apps are coupled with the use of signal strength to label devices in proximity. In this case, the devices should be able to connect over Bluetooth and the signal strength should be above a threshold i.e. the device is not just in the hearing range but also close enough to deliver a strong signal, to be identified as a contact. 
\par
Bluetooth coupled with signal strength is the most common approach out there today and it forms the bedrock of the Apple-Google Exposure Notification system \cite{Apple_Google}. However, as the founders of Bluetooth have themselves pointed out \cite{interceptBLE}, this approach is bound to have false-positives and false-negatives as well. The premise of this approach is that the signal strength is a function of distance, therefore signal strength can be used as a proxy measure for distance. Unfortunately, the signal strength depends on other factors too - such as phone context that is reflected by the phone being carried in hand, bag, clipper, or if there is an obstacle between the two devices. This leads to large errors in distance estimation where two devices 3 feet away can appear to be 30 feet away and vice-versa. Such inaccuracies in Bluetooth positioning are common knowledge among researchers in the wireless positioning community. However, in the absence of more advanced techniques that are as ubiquitous as Bluetooth, this method has emerged as the method of choice.
Governments across various countries have rolled out Bluetooth based apps - AarogyaSetu in India \cite{aarogya_setu}, TraceTogether in Singapore \cite{trace-together}, SwissCovid in Switzerland \cite{SwissCovid}, and Australia \cite{COVIDSAFE} mainly due to the ubiquity of the smartphones on which these apps are deployed.

\vspace{6pt}\noindent\textbf{GPS - } The Global Positioning System (GPS) provides geographic location and time information to devices equipped with a GPS receiver through a set of GPS radio navigation satellites. GPS provides a means to track every smartphone and identify two phones as being in contact when they stay too close for too long. This premise is promising because most smartphones are GPS-enabled and GPS works across the globe. So, a GPS-based solution would be ubiquitous and scalable. Solutions such as CovTracer \cite{CovTracer} used in Cyprus collect GPS data to compute the proximity of devices.

However, this approach fails on two counts. Firstly, it raises serious privacy questions (\emph{"Am I Being Tracked?"}) because the location (not just a contact) is being logged, and secondly, it fails on the accuracy front as well. GPS suffers from poor accuracy, especially indoors and outdoor urban tunnels. It is non-trivial to enable robust GPS indoors and even when it works, it achieves positioning errors of around 10 meters (nearly 30 feet) \cite{indoorGPS}. These errors are too high for contact-tracing approaches. A contact tracing approach needs to identify if two people are closer than 6 feet (2 meters) of each other. An error of 30 feet will throw these estimates way off and will lead to a high false-positive or false-negative rate, depending on how the system is designed.

Furthermore, GPS traces are easy to de-anonymize and can reveal a lot about individuals such as their daily routines, activities, and frequent co-locators. Contact tracing advocates believe that these concerns are overshadowed by the health benefits of the contact tracing apps. Moreover, they claim that users share their location data with corporations every day when they check their apps for maps, ride-hailing, or food delivery. However,  a user has the choice to not use GPS based apps for maps, ride-hailing, and other services due to privacy violations. But, making a GPS based contact tracing app mandatory, as is being done with contact tracing apps in some jurisdictions, will make users have no control or option to opt-out resulting in the possibility of privacy violations and attacks from corporations as well as government organizations.

Overall, the accuracy of GPS is low for contact tracing with errors of 20-30 feet, especially indoors where humans spend a lot of their time, so using it for contact-tracing won't help much anyway. Also, sharing location details with a central entity can lead to leakage and data misuse. However, with explicit user consent to upload selective data, it can provide other benefits like help governments visualize the locations of high-risk individuals, and allocate healthcare resources, thereby preempting the outbreak.

\vspace{6pt}\noindent\textbf{WiFi - } While GPS works outdoors as explained above, it has limitations indoors and in urban tunnels. People spend about 80\% of their time indoors \cite{klepeis2001national}, so we need a solution for indoor contact tracing. In homes and enterprises, WiFi networks provide an alternative approach. In enterprise WiFi networks comprising of several access points (AP), user devices constantly get connected, disconnected, and move between APs.  All these events of connections, disconnections, authorizations, and hops across APs are logged internally by each AP in a system log aka “syslog” file. By analyzing the events logged in the syslog file for each device we can identify the mobility trajectory of each device. Today's users carry their phones along everywhere and thus, the mobility trajectory of a smartphone can be considered as a proxy for the user mobility trajectory. Thus, in an enterprise network, we can derive the mobility trajectories of all users connected to the network using the WiFi logs. 

The derived mobility trajectories provide details about locations visited along with the time of visit as logged by the connection, disconnection time by the APs in the syslog files. Without advanced localization techniques like channel state based approaches, the spatial granularity is restricted to the AP range (tens of meters). While performing localization and fingerprinting leads to a finer spatial granularity for proximity analysis, it requires more effort to deploy. Once the spatio-temporal mobility trajectories of all users are derived a proximity graph also called a contact graph can be generated. If a user reports illness then the contact graph is used to identify the co-located users with the infected person over the past $n$ days, where $n$ varies based on the disease incubation period. We can use the generated contact graph to identify the set of most frequently visited locations for frequent sanitization and the set of users who visited a location after the infected person departed but during the window when the pathogens dispersed might be alive.

In \cite{trivedi2020wifitrace} authors propose a network-centric system that uses WiFi syslogs where as systems such as TraceFi \cite{TraceFi} use WiFi signals and strength to determine proximity. The main advantages of such WiFi-based contact tracing are (i) provisioning for indoor contact tracing (ii) no need for installation of new sensors or new instrumentation, works with existing WiFi network and logs collected by most enterprise for network security and traffic analysis (iii) is completely passive (iv) provides details of the time of visit and duration of the visit (v) no app installation on mobile phones needed for WiFi syslog based systems.

The main problems with WiFi-based contact tracing are (i) spatial location granularity depends on the AP range (ii) proximity v/s contact: a person might be connected to the same AP but be in different rooms. WiFi-based contact tracing might state that the 2 persons are in contact when in reality they might have never come in contact (iii) different devices have different aggressiveness to hop APs.

\vspace{6pt}\noindent\textbf{Acoustic-ranging - } Sound is as ubiquitous as Bluetooth. After all, the primary purpose of phones is to emit, record, and transfer sound (or it used to be). Thus, sound-based range estimation is another competing approach. Sound-based contact tracing is based on the idea that every device emits a random but unique sound signature that can be used to infer distance. The acoustic frequency and amplitude are chosen to be outside of the human hearing range. Unlike Bluetooth, GPS, and WiFi, sound is a mechanical wave that travels much slower than radio waves (in the order of $10^6$ times slower). This gives acoustic-ranging an arrow in its quiver, which is time-of-flight estimation. Specifically, two devices can accurately measure the time it takes for a sound to go from one device to another. Since sound travels as a wave, we can multiply the time it takes for sound to travel between devices by the speed of sound to get the distance between the devices. For example, if it took a sound signal 4 milliseconds to go from $Phone_1$ to $Phone_2$, it indicates that the phones are at a distance of about 4.5 feet. Also, an error of a millisecond or two in measuring the time of travel wouldn’t make the errors large. More importantly, the time estimates are more robust to reflections and obstacles than signal strength estimates.

Though acoustic-based systems have the pros of high accuracy and ubiquity \cite{BeepBeep}, there are 3 main pitfalls in sound-based ranging :
\begin{itemize}
\item \textit{Scale:} The number of devices that can simultaneously perform sound-based contact tracing is small. It works fine when communication is between 2 devices, but as the number of devices communicating with each other increases, it becomes very noisy. Ambient noise in commercial places such as a retail store adds further noise making the system unreliable. 
\item \textit{Privacy:} Implementing an acoustic based system using smartphones is the best solution for ubiquity. This in turn means the microphones on smartphones need to be turned \emph{“on”} all the time or sample at a high frequency raising a high level of privacy-risk that most people are unwilling to accommodate.
\item \textit{Animal Discomfort:} Though the acoustic range signals are inaudible to the human ear as they fall in the frequency range outside the human hearing range, animals can hear them. We need rigorous tests to ensure that acoustic range systems do not cause discomfort to animals that can pick them up. 
\end{itemize}

\begin{table*}
\begin{tabular}{wl{0.081\textwidth}wl{0.25\linewidth}wl{0.3\textwidth}wl{0.1\textwidth}wl{0.1\textwidth}}
\hline
Technology & Proximity Accuracy & Privacy & Ubiquity & Scalability \\ \hline
Bluetooth & Low accuracy(errors 10-20 feet). 
& Privacy preserving solution & Yes & Yes \\
GPS & Low accuracy (errors $~30$ feet) & Privacy concerns for absolute location  & Yes & Yes \\
Acoustic & High accuracy & Privacy concerns  for microphone access & Yes & No \\
WiFi Logs & Low Accuracy (errors 10-15 meters) & Privacy preserving solution & Yes & Yes \\
Advanced  & High accuracy (0-5 feet) & Privacy concerns exist & No & No \\ \hline 
\end{tabular}
\caption{Comparison of various digital contact tracing technologies}
\label{tab:comp_table}
\end{table*}

\vspace{6pt}\noindent\textbf{Non-mainstream Methods - } GPS, Bluetooth, and WiFi are relatively mainstream methods but there are several other interesting ideas and research on technology to support digital contact tracing. Below we list a few of them:
\begin{itemize}
\item \textit{Time-of-flight using Bluetooth:} Using the concept of time-of-flight measurements in Bluetooth as done in sound waves is a very interesting premise. Using Bluetooth for time-of-flight measurements will alleviate the concerns about sound-based ranging and inaccuracies of Bluetooth contact-tracing. Since radio waves travel a lot faster ($10^6$ times faster) than sound waves the time measurements will need to be highly accurate. In \cite{Chronos} and \cite{BLoc} authors present work on achieving nanosecond-level accuracy in time measurements and achieve a foot-level accuracy in distance measurements by using advanced phase information. However, phase information isn't available on all smartphones today, so this approach, while way more accurate, won't be ubiquitous for the next few years.
\item \textit{Hybrid multi-modal approaches:} Many app designers today are exploring the use of different modalities like sound and Bluetooth in conjunction with each other so that one technology can help remove the pitfalls of the other. Such approaches are seeing a lot of traction and are being used in many of the apps out there. These approaches, however, are still in the preliminary stages of research and haven't been tested extensively.
\end{itemize}

\par
Table \ref{tab:comp_table} compares these technologies across the parameters of accuracy, privacy, ubiquity, and scalability.

\subsection{Choice of digital contact tracing solutions}
All digital contact tracing systems can be broadly classified into systems that collect absolute location data (eg. GPS, WiFi logs, etc) and relative location data (Bluetooth). Based on this classification, the system functionality itself is different. Systems that collect absolute location data can also find out same place different time contacts in addition to same place same time contacts. Such systems are very useful in contact tracing of infectious diseases that spread indirectly. When a patient visits a location and spreads the pathogens at a location, the pathogens remain active at the location for a certain time duration. Even though the patient leaves the location, a visitor may get infected if he/she visits the same location and comes in contact with surfaces with active pathogens resulting in indirect transmission of the disease.

On the other hand, systems that rely on relative location of two devices, i.e. they can tell if person A is close to person B but not where they are, offer exciting features from a privacy standpoint. The inherent guarantees are stronger, even though you can still reveal information about who interacts with whom. Though a lot of features of the contact tracing system are defined by the above classification, the choice of contact tracing system also depends on several other factors discussed below :
\begin{itemize}
\item \textbf{Location Granularity -} Location granularity defines the range or reach within which two or more people are labeled in the proximity of each other. The coarser the location granularity more people are reported to be in proximity and the lower is the system accuracy in identifying true co-locators. For eg, a Call Data Record based contact tracing system reports all users within the range of the call tower to be colocated. The range of the call tower can range from 15 Kms to 50 Kms whereas a Bluetooth based system has a location granularity of 10 meters and hence reports a lower number of people who are truly within the proximity range and has lower false positives of labeling co-locators than a coarse location-based system. GPS based systems have finer location granularity, Bluetooth based systems have a location granularity of 10 meters, WiFi-based systems (without advanced signal processing techniques) have a location granularity based on the WiFi range of the access point.
\item \textbf{Temporal Granularity -} Temporal granularity provides the user time of visit at a location. This helps identify if two people are co-located at the same time for direct transmission or visited the same location within the time frame the pathogens were active. In the case of WiFi-based systems, the temporal granularity is also based on the aggressiveness of the device in hopping across APs.
\item \textbf{Indoor v/s Outdoor -} Some technologies such as GPS only work outdoors while WiFi, location beacons, Bluetooth, Acoustic Range systems can all work indoors. 
\item \textbf{Percentage Population Adaption-} Not everyone owns a smartphone with the latest technological advancements that could result in a low percentage of user population adapting the digital contact tracing solution. Some estimates \cite{WSJ-TT} suggest an adaptation of 60\% - 80\% of the population to use digital contact tracing to add value. This is problematic in vulnerable populations like prison inmates, poverty-stricken neighborhoods, etc. where the smartphone density is low. If digital contact-tracing alone is used for allocating healthcare resources and become the center of our strategies, we will end up completely missing out on large groups of people.
\item \textbf{Scalability -} Understanding the user population size for performing contact tracing is an important parameter in selecting the technology. The system should be scalable to support a large enough population while providing high availability to prove efficient. 
\item \textbf{Data Storage -} The collected data can be stored in 2 ways either all the collected data gets pooled into a central data store or is distributed across multiple data stores to create a distributed system. Understanding the limitations of the environment and the features of the system would help decide the data storage architecture aspect of the digital contact tracing system. 
\item \textbf{Data Collection Reliability -} The underlying user location dataset needed for the effective working of the contact tracing system needs to be collected reliably, highly available, and representative of a sufficiently large population for effective contact tracing. If the collection mode requires the consent of users then a substantial population of the users should use the app, give consent to data collection and a reliable mode of uploading the data to a system for analysis should exist. In contrast, if a system is designed to use passive data from sources already collecting the data for other purposes such as network analysis for performance or security such as WiFi logs, then only the consent from the user for health data needs to be taken.
\end{itemize}


\section{Security and Privacy}
\label{sec:sec_privacy}

Any digital contact tracing tool requires the collection of user location/proximity data (Spatio-temporal component), and user health data forming the base dataset for digital contact tracing. 
The channels of data collection depend on the technology used such as GPS logs comprising of latitude, longitude details for GPS based solutions, WiFi syslog for WiFi based solutions, Bluetooth beacons, and user ids for BlueTooth based proximity identification, etc. The storing mechanisms are based on technology as well as system design and fall under 2 main categories: (i) centralized datastore resulting in a single contact tracing graph, and (ii) distributed datastore. Any method of data collection and storage has 2 main user concerns linked to it - Privacy and Security. In this section, we briefly discuss the main aspects of privacy and security.

\vspace{6pt}\noindent\textbf{Security: }
As stated above, all contact tracing systems deal with extremely sensitive data due to the collection and storage of user personal data. This makes the system susceptible to security exploits and attacks. A secure system should satisfy the three main areas namely confidentiality, integrity, and availability. These three aspects of security should protect the user data from identification, alteration, access by non-privileged users, disclosure, data safety, and authentication. Authenticity is an important criterion and guarantees that the user location data (absolute or relative location data) is not forged. Strict rules should be defined for data security and user information confidentiality. Apart from confidential, password protected data storage and access mechanisms to the system code should be well subjected to thorough security audits to ensure protection against attacks.

\vspace{6pt}\noindent\textbf{Privacy: }
Designing an effective privacy preserving contact tracing system is an essential requirement in making the users feel safe and secure enough to voluntarily opt-in and use a solution. First, note that contact tracing itself (digital or not) reveals some private information about users -- their health status and their interactions with other individuals. In some contexts, access to this information can be harmful, for example, when interactions between government critics are revealed in authoritarian regimes or when individuals get targeted for their visits to LGBTQ meetings. For a digital contact tracing solution, the goal is to (a) not reveal or store more information than is absolutely required, and (b) to prevent such misuse of the data.

Recent research interest in privacy preserving contact tracing solutions has been explored leading to some interesting solutions. Researchers are exploring various protocols,  and mechanisms for designing a privacy preserving digital contact tracing solutions \cite{chan2020pact, liu2020privacy, Trieu2020EpioneLC, Verssimo2020PriLokCD}. For example, the generation of random unique identifiers for devices while using Bluetooth for finding proximity. In such solutions \cite{trace-together}, the phone is assigned random identifiers and so are other smartphones in proximity. The phone keeps a track of its unique identifiers as well as unique identifiers of the other phone in proximity as broadcast by them. When a particular smartphone user is marked as an infected person then the phone identifier labels of the infected person are announced publicly and each user checks the infected user phone identifiers against the self log of proximity phone identifiers to check if a proximity contact was ever established. This approach keeps the user data private.

The challenge in this approach is the same as that of Bluetooth based apps resulting in high false positives, which could result in the system becoming unusable while creating panic. To avoid this problem the signal strength associated with each message can be used to compute the proximity distance between the devices. The weaker the signal strength the farther away is the device and vice versa. However, quantifying the signal strength value to classify devices as close and far is tricky and depends on the model, make of phone, phone context (in the pocket, hand, bag, phone case, etc).

To concur, in the case of contact tracing systems, since access to user health data and location data both are needed for the correct functioning of the system it becomes extremely crucial that the system specifications, design, and development be considered as inter-disciplinary user-focus research taking into account the social, psychological, security and Human-Computer Interaction (HCI) aspects of users while designing solutions satisfying the needs of the healthcare workers.

\section{Path Forward and Conclusions}
\label{sec:conclusions}

In this paper, we presented various technologies that can be used for contact tracing and discussed various factors that play important role in making the design decisions of a contact tracing system. While digital contact tracing has a lot of benefits especially with respect to scalability, quick reach, and precision there are pressing issues concerning privacy and security that should not make the system feel and act like a surveillance system. 
User data leakage and use by agencies such as government that could result in the exploitation of citizens should never happen. We need to solve these challenges with technological as well as legal solutions through better privacy laws. Additionally, we need to design frameworks that are compatible with contact tracing systems that use different channels of data collection. We strongly believe that designing an efficient privacy preserving contact tracing system should be an inter-disciplinary  user-focus  research  taking into account the social, psychological, security, and Human-Computer Interaction aspects of users while designing solutions satisfying the needs of the healthcare workers.

\begin{acks}
Amee would like to thank John Krasinski and Sendhil Ramamurthy for helping her stay sane during the pandemic. Deepak thanks Ranveer Chandra for seeding some of the ideas in this editorial.
\end{acks}

{ \balance
{
\def\bibsection{\section*{References}}
	\bibliographystyle{ACM-Reference-Format}
    
    \bibliography{main}
    	

        
}
}


\end{document}